# Post-Nonlinear Sparse Component Analysis Using Single-Source Zones and Functional Data Clustering

Matthieu Puigt, Anthony Griffin, and Athanasios Mouchtaris,

*Abstract*—In this paper, we introduce a general extension of linear sparse component analysis (SCA) approaches to postnonlinear (PNL) mixtures. In particular, and contrary to the state-of-art methods, our approaches use a weak sparsity source assumption: we look for tiny temporal zones where only one source is active. We investigate two nonlinear single-source confidence measures, using the mutual information and a local linear tangent space approximation (LTSA). For this latter measure, we derive two extensions of linear single-source measures, respectively based on correlation (LTSA-correlation) and eigenvalues (LTSA-PCA). A second novelty of our approach consists of applying functional data clustering techniques to the scattered observations in the above single-source zones, thus allowing us to accurately estimate them. We first study a classical approach using a B-spline approximation, and then two approaches which locally approximate the nonlinear functions as lines. Finally, we extend our PNL methods to more general nonlinear mixtures. Combining single-source zones and functional data clustering allows us to tackle speech signals, which has never been performed by other PNL-SCA methods. We investigate the performance of our approaches with simulated PNL mixtures of real speech signals. Both the mutual information and the LTSA-correlation measures are better-suited to detecting single-source zones than the LTSA-PCA measure. We also find local-linear-approximation-based clustering approaches to be more flexible and more accurate than the B-spline one.

## I. INTRODUCTION

**B**LIND Source Separation (BSS) consists of estimating a set of $N$ unknown source signals $s_j$ from a set of $P$ observations $x_i$ resulting from mixtures of these sources through unknown propagation channels [1]. Among all the proposed approaches, the ones based on sources joint-sparsity, known under the name of Sparse Component Analysis (SCA) methods, have met with great interest in the community in the last decade (see e.g. [1, Ch. 10]). Indeed, they are naturally adapted to stationary, non-stationary and/or dependent signals and are thus an alternative to classical Independent Component Analysis (ICA) approaches which assume source mutual independence. Moreover, they allow processing of the underdetermined case where $N > P$.

Most of the SCA approaches have been proposed for linear mixtures, i.e. linear instantaneous (LI), anechoic or convolutive mixtures. While many methods assume the sources to be (approximately) W-disjoint orthogonal (WDO) in an analysis domain[1] [2], several other methods highly relax this assumption, by looking for "single-source zones" (i.e. zones where one source is dominant over the others) [3]–[7]. Interestingly, while many SCA methods have been proposed for linear mixtures, only a few sparsity-based methods process nonlinear configurations [8]–[11]. In [8], [9], the authors consider postnonlinear (PNL) mixtures (i.e. a special configuration where linear mixes of sources are distorted by a function which models data acquisition/sensor nonlinearities, such as saturation), and assume the sources to be approximately WDO[2]. Unfortunately, these approaches are not tested with real-life source signals, mainly because of the strong sparsity assumption. In [10], [11], the authors extend the measures for finding single-source zones to other classes of nonlinear mixtures but restrict their approach to overdetermined or determined mixtures.

In this paper, we propose an approach for identifying PNL mixtures[3] based on single-source zones, as in [10], [11], and which possibly processes the underdetermined case, as in [8], [9]. We thus avoid the strong source sparsity assumption of [8], [9] while processing the same class of mixtures and applying our approach to mixtures of real speech signals. Our main contribution is dedicated to the estimation of nonlinear mappings, by combining single-source zones (found using confidence measures well-suited to nonlinear mixtures) and functional data clustering. We thus provide a way to extend linear SCA [3]–[7] to PNL-SCA. This work has been partially proposed in [12]. However, here we extend [12] in several ways: we propose several single-source confidence measures well-suited to PNL mixtures and several methods to cluster the functional data points. Moreover, we present an exhaustive experimental validation of the approaches. An extension of the proposed approaches to more general mixtures, partially proposed in [13], is also investigated in this paper. In particular, here we give a better characterization of the achieved performance than in [13].

The remainder of the paper is structured as follows: in Section II, we describe the considered BSS problem. We then introduce our proposed method in Sections III, IV, and V. Section VI provides an experimental validation of the approach and we conclude and discuss future work in Section VII. Appendix A introduces the extensions of the proposed PNL

M. Puigt and A. Griffin are with the Institute of Computer Science of the Foundation for Research and Technology – Hellas (FORTH-ICS), Heraklion, Crete, Greece.

A. Mouchtaris is with FORTH-ICS and the Department of Computer Science of the University of Crete, Heraklion, Crete, Greece.

---

[1]The WDO assumption means that in each atom of an analysis domain (e.g. time, time-frequency, time-scale domain), at most one source is non-zero.

[2]Actually, in [8], the authors assume the sources to be $(P-1)$-sparse, which is equivalent to WDO if $P = 2$. In [9], the approximate WDO is not explicitly assumed but is needed by authors and satisfied in their tests.

[3]An extension to other nonlinear mixtures is provided in Appendix A.

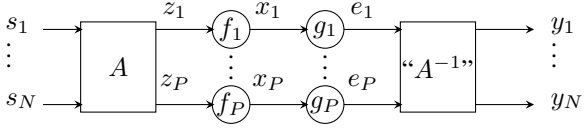

Fig. 1. PNL Mixing-separating structure.

approaches to more general nonlinear mappings.

## II. PROBLEM STATEMENT, DEFINITIONS AND ASSUMPTIONS

In this paper, we assume that $N$ real source signals $\underline{s}(t) = [s_1(t), \ldots, s_N(t)]^T$ are mixed by an LI unknown $P \times N$ mixing matrix $A$, thus providing a set of linearly mixed signals

$$\underline{z}(t) = A\underline{s}(t), \quad (1)$$

to which a nonlinear componentwise vector mapping $\underline{f} = [f_1, \ldots, f_P]^T$, assumed to be invertible, is applied. It can e.g. model data acquisition/sensor nonlinearities such as saturation. Such a situation e.g. arises in audio processing with small and cheap microphones in a mobile device. Observed signals $\underline{x}(t)$ thus read

$$\underline{x}(t) = \underline{f}(\underline{z}(t)) = \underline{f}(A\underline{s}(t)). \quad (2)$$

We aim to estimate the source signals $\underline{s}(t)$, up to a scale coefficient/permutation indeterminacy. This means that we want to suppress or minimize the distortions introduced by the nonlinear mappings $f_i$. For that purpose, we use a separating structure which is the *mirror* of the mixing one (see e.g. [1, Ch. 14]): we first have to estimate $g_i$, the inverse of the nonlinear mappings $f_i$, and to apply them to the observations. We then obtain a linear problem comparable to (1) and we process a linear SCA approach to estimate the sources. The global mixing and separating structure is shown in Fig. 1. The proposed separating structure may be summarized as follows:

1) We first look for temporal zones where one source is dominant over the others (see Section III).
2) We then estimate the nonlinear mappings $f_i$ (see Section IV).
3) We then invert the nonlinearities and get an LI-BSS problem, that we solve using an LI-SCA approach (see Section V).

Before introducing the proposed approach, we first introduce the only assumptions of the proposed approach and their associated definitions.

*Definition 1 ( [8]):* Let $A = [a_{ij}]$ be an $P \times N$ matrix. Then $A$ is said to be "mixing" if $A$ has at least two nonzero entries in each row. And $A$ is said to be "absolutely degenerate" if there are two normalised columns $a_{ik}$ and $a_{il}$ with $k \neq l$ such that $|a_{ik}| = |a_{il}|$, i.e. $a_{ik}$ and $a_{il}$ differ only by the sign of the entries.

*Assumption 1 (Mixing assumptions):*

1) $A$ has nonzero entries on the first row and at least one nonzero entry in each other row. We cannot find two collinear vectors $|a_{ik}|$ and $|a_{il}|$, with $k \neq l$.

2) In the underdetermined case when $N > P$, every $P \times P$-submatrix of $A$ is invertible.
3) We also assume that, for each index $i$, $f_i(0) = 0$.

Assumption 1.1 is needed for the following reasons [8]: if $A$ is not "mixing" (according to Definition 1), then this means that there is an index $i$ such that the $i$-th row of $A$ contains only one non-zero element and consequently $f_i$ cannot be identified. As an extreme case, let us imagine that $A$ is diagonal (up to a permutation in the order of its columns), then each observed signal reads

$$x_i(t) = f_i(a_{ik}s_k(t)), \quad (3)$$

i.e. we already get separated signals with which we can do nothing more without extra information. If $A$ is absolutely degenerate, it can be estimated, but the nonlinear mappings cannot [8]. Assumption 1.2 is a classical assumption in underdetermined BSS. This means that locally, if only $P$ sources are active, we get a determined BSS problem which needs to be separable. Lastly, Assumption 1.3 is not limiting for practical applications and is shared by many PNL-BSS approaches.

*Definition 2:* A "temporal analysis zone" is a subset $\mathcal{T}$ of the time domain.

From a theoretical point of view, each temporal analysis zone may be set to any kind of subset of the time domain, and might even be restricted to a single time instance $t$. However, in practice, we set these zones to temporal intervals, denoted $\mathcal{T}$.

*Definition 3:* A source is said to be "isolated" in a temporal analysis zone $\mathcal{T}$ if only this source (among all considered mixed sources) has a nonzero variance in this zone. We then say that this zone is "single-source".

This definition corresponds to the theoretical point of view. From a practical point of view, this means that the energy of all other sources is negligible with respect to the energy of the source which is isolated.

*Definition 4:* A source is said to be "accessible" in the time domain if there exists at least one temporal analysis zone where it is isolated.

*Assumption 2 (Source assumptions):*

1) Source signals are mutually independent.
2) At least $P$ sources are accessible in the time domain.
3) By considering *several* single-source zones associated with the same source, the amplitude of the observations spans a "wide" range allowing the estimation of the nonlinear functions $f_i$.

Note that, contrary to linear BSS methods [3], [6] which needed source linear independence, here we need source mutual independence. This is due to the more complex mixing model, as we will see in Section III. We need $P$ isolated sources in order to be able to invert the $P$ nonlinear functions $f_i$ [8]. Assumption 2.3 is needed because we want to estimate the nonlinear mappings $f_i$ on their whole domains. In the case where we would be able to estimate the functions $f_i$ on a subset of their domain, the whole estimation might be coarse, thus yielding a poor quality of separation.

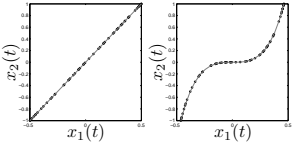

Fig. 2. Scatter plots of observations on a single-source zone and theoretical curves. *Left:* LI mixtures. *Right:* PNL mixtures.

## III. NONLINEAR SINGLE-SOURCE CONFIDENCE MEASURES

Before introducing the proposed method, let us focus on the intuitive idea behind it. If a source, say $s_k$ is isolated, then

$$x_i(t) = f_i(a_{ik} s_k(t)), \quad \forall i \in \{1, \ldots, P\}, \forall t \in \mathcal{T}, \quad (4)$$

and we obtain, assuming that $a_{ik} \neq 0$,

$$s_k(t) = \frac{f_i^{-1}(x_i(t))}{a_{ik}}, \quad \forall i \in \{1, \ldots, P\}, \forall t \in \mathcal{T}. \quad (5)$$

We thus have the following relationship between observations $x_1$ and $x_i$ for any $t \in \mathcal{T}$:

$$x_i(t) = f_i\left(\frac{a_{ik}}{a_{1k}} f_1^{-1}(x_1(t))\right) = \phi_{ik}(x_1(t)), \quad (6)$$

where the functions $\phi_{ik}$ are defined as:

$$\phi_{ik}(u) = f_i\left(\frac{a_{ik}}{a_{1k}} f_1^{-1}(u)\right). \quad (7)$$

The right plot of Fig. 2 shows an example with $P = 2$ observations of the behaviour of data points $\underline{x}(t)$ in a single-source zone.

In an LI problem, the relationships (6) between observations are much simpler. Let us first recall that in this case, the mapping $\underline{f}$ in Eq. (2) is set to the identity function, i.e. the observations $\underline{x}(t)$ are equal to the signals $\underline{z}(t)$ defined in Eq. (1). In the case when a source, say $s_k$ is isolated, then observations read:

$$x_i(t) = a_{ik} s_k(t), \quad \forall i \in \{1, \ldots, P\}, \forall t \in \mathcal{T}, \quad (8)$$

and we get the following relationship between observations $x_1$ and $x_i$ (see the left plot of Fig. 2 for an example with $P = 2$ observations):

$$a_{ik} x_1(t) - a_{1k} x_i(t) = 0, \quad \forall i \in \{1, \ldots, P\}, \forall t \in \mathcal{T}. \quad (9)$$

In [3]–[7], authors proposed finding such single-source zones by means of a "single-source confidence measure" based on asymmetrical [3] or symmetrical [4] ratios of observations, correlation [5], [6], and local PCA [7]. In this paper, we need to extend such measures to the considered PNL mixture, where there is a functional relationship between observations.

### A. Mutual information as a nonlinear correlation measure

We explained above how authors provided ways to find single-source zones when linear mixtures are applied to the sources. For example, in the case where we estimate the correlation coefficient of a pair of observations [5], [6], this coefficient is equal to 1 in absolute value when one source is isolated and is much lower otherwise. In the considered PNL mixture, one needs to measure a *nonlinear* correlation between observations. Mutual Information $\mathcal{I}(\underline{x})$, defined as:

$$\mathcal{I}(\underline{x}) = -\mathbb{E}\left\{\log \frac{\prod_{i=1}^{P} \mathbb{P}_{x_i}(x_i)}{\mathbb{P}_{\underline{x}}(\underline{x})}\right\}, \quad (10)$$

where $\mathbb{E}\{.\}$ stands for expectation, $\mathbb{P}_{\underline{x}}$ and $\mathbb{P}_{x_i}$ ($i \in \{1, \ldots, P\}$) are the joint and marginal probability density functions of the observations respectively, provides such a measure [14]: it takes null values when variables are independent and much higher values otherwise. However, if we want this measure to have the same behavior as linear correlation, we need to normalize it, which is classically done as follows:

$$\mathcal{I}_{\text{norm}}(\underline{x}) = \sqrt{1 - e^{-2\mathcal{I}(\underline{x})}}. \quad (11)$$

This measure has also been used in [11] for another class of nonlinear mixtures, and we use it in the same way as [11]: a source is isolated in an analysis zone iff $\mathcal{I}_{\text{norm}}(\underline{x}) = 1$. We thus only consider the analysis zones which maximize Eq. (11).

### B. Manifold learning based measures

In contrast to mutual information, as we assume the NL mappings $f_i$ to be smooth, the resulting functions $\phi_{ik}$ are also smooth and one may *locally* consider them as linear. Such an idea is quite common in manifold learning [15] and allows us to extend linear single-source approaches to PNL configurations.

For example, the Local Tangent Space Approximation (LTSA) [15, Sect. 3.2.4] method learns the manifold by constructing the local tangent space of each observed data point. We propose using such an idea to extend linear single-source confidence measures [3]–[7] to PNL mixtures. Our approach consists of successively considering each sample $\underline{x}(t_i)$ of the analysis zone $\mathcal{T}$, defining its neighborhood (estimated by means of a $K$-nearest neighbor ($K$-NN) technique), and applying a linear single-source measure in this neighborhood.

As in [7], we realize an eigendecomposition of the correlation matrix of the data. If one source is isolated, then the rank of the observations is equal to 1 and the highest eigenvalue $\lambda_1(t_i)$ is non-negligible while the $K-1$ other ones $\lambda_j(t_i)$ are close to zero. The authors of [7] proposed computing the ratio between the highest eigenvalue and the sum of the others to find single-source zones. In this paper—in order to keep an analogy with the behavior of the correlation—we propose computing the ratio[4]:

$$R(t_i) = \frac{\lambda_1(t_i)}{\sum_{j=1}^{K} \lambda_j(t_i)}, \quad (12)$$

which is close to 1 in single-source zones and much lower otherwise. Once we compute the ratios $R(t_i)$ for all the data points observed at time $t_i \in \mathcal{T}$, we derive the actual *global* single-source confidence measure, denoted $\mathcal{R}(\underline{x})$ hereafter, as

---

[4]Note that such idea has been proposed in [16] for selecting "simple autoterms" of bilinear time-frequency transforms.

the geometric mean of all these ratios:

$$\mathcal{R}(\underline{x}) \triangleq \left(\prod_{t_i \in \mathcal{T}} R(t_i)\right)^{\frac{1}{|\mathcal{T}|}}, \quad (13)$$

where $|\mathcal{T}|$ is the cardinality of the zone $\mathcal{T}$, i.e. the number of time samples it contains. The extension of other single-source measures using this framework is straightforward. In [5] and [6], the authors propose computing the covariance or the correlation coefficient between observations. Here, we compute the correlation coefficient between observations 1 and $j$, in the neighborhood of observed point $\underline{x}(t_i)$ at time $t_i$:

$$C_{x_1, x_j}(t_i) = \frac{\mathbb{E}\{x_1(t_i) x_j(t_i)\}}{\sqrt{\mathbb{E}\{x_1(t_i) x_1(t_i)\} \mathbb{E}\{x_j(t_i) x_j(t_i)\}}}, \quad (14)$$

we average this over the indices of $j$ (we then denote it $C_{\underline{x}}(t_i)$) and we derive the global single-source confidence measure which is well-suited to PNL mixtures, denoted by $\mathcal{C}(\underline{x})$ hereafter:

$$\mathcal{C}(\underline{x}) \triangleq \left(\prod_{t_i \in \mathcal{T}} C_{\underline{x}}(t_i)\right)^{\frac{1}{|\mathcal{T}|}}. \quad (15)$$

Note that other linear single-source confidence measures might be extended as well, as in the work in [3] and [4] for example.

*C. Constant sources and extended single-source confidence measures*

So far, we have proposed single-source confidence measures assuming that the energy of the absent sources is very low. A problem may appear if, in a zone $\mathcal{T}$, sources $s_j(t)$ are constant but not null[5]. We denote these constant values by $\overline{s_j}$. In this situation, we still have our above single-source confidence measures equal to 1 but Eq. (4) then becomes

$$x_i(t) = f_i \left(a_{ik} s_k(t) + \alpha_{i,k}(\mathcal{T})\right), \quad \forall i \in \{1, \ldots, P\}, \forall t \in \mathcal{T} \quad (16)$$

where $\alpha_{i,k}(\mathcal{T}) = \sum_{j \neq k} a_{ij} \overline{s_j}$. Eq. (5) and (6) then resp. read

$$s_k(t) = \frac{f_i^{-1}(x_i(t)) - \alpha_{i,k}(\mathcal{T})}{a_{ik}}, \quad \forall i \in \{1, \ldots, P\}, \quad (17)$$

and

$$x_i(t) = f_i \left(\frac{a_{ik}}{a_{1k}} \left(f_1^{-1}(x_1(t))\right) - \frac{a_{ik}}{a_{1k}} \alpha_{1,k}(\mathcal{T}) + \alpha_{i,k}(\mathcal{T})\right). \quad (18)$$

Let us recall that we are looking for zones where all the constant coefficients $\alpha_{i,k}(\mathcal{T})$ are zero. As we are applying the proposed approach to speech signals, the situation when one can find an index $i$ such that $\alpha_{i,k}(\mathcal{T}) \neq 0$ will not occur. Additionally, due to Assumption 1.3, we know the value of each nonlinear function $f_i$ is zero at zero and we can estimate

---

[5]Such a scenario is not a problem in LI-SCA: observations can be locally centered in each analysis zone, thus zeroing the constant signals [5]. Moreover, such constant sources provide the main difference between our proposed method and [11]: we are looking for zones where all the sources are zero except one while the authors of [11] are looking for zones where all the sources are strictly positive constant except one.

---

$\widehat{\phi_{ik}}$, the nonlinear relationship between observations defined in Eq. (7) (see Section IV) and discard the zones where $\widehat{\phi_{ik}}(0) \neq 0$.

Finally, from a theoretical point of view, we should look for analysis zones which satisfy:

$$\begin{cases} \text{SSCM}(\underline{x}) & = & 1 \\ \widehat{\phi_{ik}}(0) & = & 0 \quad \forall i \in \{2, \ldots, P\}, \end{cases} \quad (19)$$

where $\text{SSCM}(\underline{x})$ is one of the single-source confidence measures defined in Eqs. (11), (13), and (15). However, in practice, we only consider zones which *approximately* satisfy the criterion (19), i.e. we look for zones $\mathcal{T}$ such that:

$$\begin{cases} \text{SSCM}(\underline{x}) & > & 1 - \epsilon_1 \\ \left|\widehat{\phi_{ik}}(0)\right| & < & \epsilon_2 \quad \forall i \in \{2, \ldots, P\} \end{cases}, \quad (20)$$

where $\epsilon_1$ and $\epsilon_2$ are some user-defined thresholds.

## IV. FUNCTIONAL DATA CLUSTERING

If we consider all the single-source analysis zones, then we get a subset of the original observations where the approximate WDO assumption holds. We can thus use the clustering techniques proposed in [8], [9]. However, in [8], the authors propose a geometrical preprocessing which is not robust to noise in general and in particular not to *non-ideal* single-source zones[6]. On the other hand, [9] proposes the use of a spectral clustering technique in order to separate the curves, and thus the sources. Spectral clustering techniques are well suited to nonlinearities in the data and are more robust to noise than the approach proposed in [8]. However, such techniques are sensitive to the distance between the curves and do not allow the clusters to overlap. This last criterion is obviously not satisfied in the BSS framework, at least around zero where all the clusters meet, and the authors of [9] proposed a solution for this last case: they remove the points of $\underline{x}$ which are close to zero, and try to find $2N$ clusters. By assuming that the nonlinear mappings are almost linear for the lowest values of $\underline{x}$, they link the half-clusters.

In this paper, we propose taking advantage of the single-source analysis zones to cluster our data. Indeed, in each single-source zone, as we saw above (see Fig. 2), all the points belong to the same functional curve and give us extra information which is not provided in [8], [9]. We can thus cluster the data according to these zones. The underlying idea can be seen as an extension of scale-coefficients clustering in [6], [7] to nonlinear mixtures: while the linear relationships between observations were limited to scale coefficients to be clustered, here we have to cluster the scattered curves observed in the single-source zones, i.e. to estimate some parameters adequately describing the functions $\phi_{ik}$ defined in Eq. (7) to realize a cluster. Such techniques are named functional data clustering, and belong to a rich topic in mathematics, named functional data analysis, which aims to study infinite dimensional data as functions [17].

Many approaches for clustering functional data belong to one of both following families. (i) The *regularization* ap-proaches consist of successively interpolating each observed scattered function, of discretizing all of them on the same time grid and lastly clustering them while considered as a high dimensional vectors. However, they are often highly correlated and may lead to unstable solutions, because of the *curse of dimensionality* [18]. (ii) The *filtering* methods consist of approximating each curve with respect to a common finite dimensional basis, and then of clustering the resulting basis coefficients.

---

[6]An ideal single-source zone is an analysis zone where all the sources except one are *exactly* zero.

A description of a simple filtering approach, using B-splines is provided in the following subsection while we derive an original clustering method from our considered PNL problem in Section IV-B.

### A. Filtering functional data clustering

Given an interval $[x_1(t_b), x_1(t_e)]$, we define a subdivision $\xi_0 = x_1(t_b) \leq \xi_1 \leq \xi_2 \leq \ldots \leq \xi_K \leq \xi_{K+1} = x_1(t_e)$. The points $\xi_l$ are named *knots*. Note that a same knot may be repeated several times, say $p$ times. We then say that it is a multiple knot of *multiplicity order* $p$. We aim to fit the curve $\{(x_1(t_j), x_i(t_j))\}_{j=1,\ldots,M}$ on such an interval by using splines. A spline is a polynomial of degree $d$ (or order $d+1$) on any interval $[\xi_{l-1}, \xi_l)$ which has $d+1$ continuous derivatives on the open interval $[x_1(t_b), x_1(t_e))$. For a fixed sequence of knots, the set of such splines is a linear space of functions with $K+d+1$ free parameters. A useful basis $(B_{1,d}, \ldots, B_{K+d+1,d})$ for this linear space is given by B-splines [20], recursively defined as

$$\begin{cases} B_{l,0}(t) &= \begin{cases} 1 & \text{if } \xi_l \leq t < \xi_{l+1} \\ 0 & \text{otherwise} \end{cases}, \\ B_{l,u}(t) &= \frac{t-\xi_l}{\xi_{l+u}-\xi_l} B_{l,u-1}(t) + \frac{\xi_{l+u+1}-t}{\xi_{l+u+1}-\xi_{l+1}} B_{l+1,u-1}(t). \end{cases} \quad (21)$$

A spline, denoted $\zeta(x_1, \underline{\beta})$ hereafter, can now be written with respect to the above basis:

$$\zeta(x_1, \underline{\beta}) = \sum_{l=1}^{K+d+1} \beta_l B_{l,d}(x_1), \quad (22)$$

where $\underline{\beta} = [\beta_1, \ldots, \beta_{K+d+1}]^T$ are the spline coefficients. For a set of fixed knots, the coefficients $\underline{\beta}$ may be estimated as a linear problem. Once the basis is estimated according to Eq. (21), the application of the B-splines is not more difficult than polynomial regression. Let $\{(x_1(t_j), x_i(t_j))\}_{j=1,\ldots,M}$ be a regression type data set of $M$ measurements of the curve $\phi_{ik}$ defined in Eq. (7). The spline coefficients are estimated as:

$$\widehat{\beta}_l = \arg\min_{\beta_l} \frac{1}{M} \sum_{j=1}^{M} \left(x_i(t_j) - \zeta(x_1(t_j), \underline{\beta})\right)^2. \quad (23)$$

Once the knots are fixed, the estimated coefficients $\widehat{\beta}_l$ describe the function shape. If we use the *same* knots for *all* the single-source analysis zones selected in Section III, then all the $K+d+1$ B-spline coefficients have the same meaning and may be compared. If two curves have close estimated B-splines coefficients, then they should be associated with the same source. Otherwise, they should be associated with different sources. Clustering techniques can be applied to such coefficients [19]: while the authors of [19] used K-means to this end, we propose to use the median-based version of K-means, named K-medians which has been used in [6]. Other approaches, such as DEMIX [7], may also be employed. Let us recall that prior to the clustering stage, we discard the single-source zones $\mathcal{T}$ which do not satisfy Eq. (19). In practice, $\widehat{\phi_{ik}}(0)$ is estimated as the value of the spline $\zeta(x_1, \underline{\beta})$ at zero.

### B. Clustering using locally-linear assumptions

In the previous section, we presented a clustering method developed for functional data. However, it suffers from the fact that finding optimal knot locations for the set of scattered data is data-dependent. Moreover, there are relationships, named Schoenberg-Whitney conditions, between the degree of the B-splines, the knot locations and the number of points between two knots to be satisfied [21].

We now propose an alternative solution based on the locally linear data approximation around zero. Indeed, in many systems where PNL mixtures hold, the NL function due to the sensor response, e.g. microphones, is almost linear around zero. The first order of the Taylor series expansion of the functions $\phi_{ik}$ reads

$$\phi_{ik}(t) = \phi_{ik}(0) + \phi'_{ik}(0) \cdot t + \mathcal{O}(t^2) \simeq \phi'_{ik}(0) \cdot t. \quad (24)$$

Eq. (24) thus reveals that the scattered functional curve is approximated by the slope of its tangent at zero and we can use this slope as a way to cluster the estimated functions. Moreover, speech signals (that we aim to process in this paper) tend to be distributed around zero in single-source zones $\mathcal{T}$ and we get a high probability for having many data points in the neighborhood of zero in each single-source zone $\mathcal{T}$. Estimating $\phi'_{ik}(0)$ may thus be done by LI-SCA, since Eq. (4) combined with Eq. (24) is equivalent to Eq.(8). As in Section III, we propose using the formalism of manifold learning, and estimating the neighborhood of 0 with the $K$-NN method, before applying one of the methods in [7], [10] to estimate the slope of $\phi'_{ik}(0)$. As explained in Section III-C, we discard single-source zones $\mathcal{T}$ which do not satisfy Eq. (19). Note that $\widehat{\phi_{ik}}(0)$ is the value at the origin of the line defined in Eq. (24). In practice, we estimate it using a least-mean square regression technique. We finally cluster the retained curves by applying a clustering method, e.g. K-medians, on these slopes.

## V. NONLINEAR INVERSION AND LINEAR IDENTIFICATION

Once the nonlinear functions are estimated, we have to invert them and apply these inverse functions to the observations $\underline{x}(t)$, in order to get the linearly mixed signals $\underline{e}(t) = [e_1(t), \ldots, e_P(t)]^T$ (see Fig. 1). This is straightforward by e.g. applying one of the neural-network-based methods, proposed in [8], [9], which use the same property but are differently implemented. The underlying common property was first defined for PNL-ICA methods [1, Ch. 14] and is adapted to PNL-SCA as follows: we estimate a nonlinear mapping $\underline{g} = [g_1, \ldots, g_P]^T$ such that for all indices $i, k$, the compound function $g_i \circ \phi_{ik}$ is linear (see Fig. 1). To this end, [8] proposes finding a linear relationship between the same components of different clusters while [9] suggests finding a linear relationship between different components in the same cluster.

Once the nonlinearities are inverted, we thus obtain a classical LI-SCA problem. The estimation of the linear mixing parameters is then straightforward if we have estimated the $N$ nonlinear curves: once we have linearized the clusters obtained in Section IV, each cluster fits a line whose parameters—defined in Eq. (9)—may be estimated using a criterion proposed in [3]–[7].

If we estimated more than $P$ curves but less than $N$, we are still able to invert the nonlinearities. However, we now do not have all the linearized curves and we will thus have to estimate the linear mixing parameters thanks to a *whole* linear SCA approach, and probably by first applying a linear sparsifying transform to $\underline{e}(t)$, in order to find a zone associated with each source.

Note that so far, we have mainly focused on the first stages of the complete approaches, which contain our proposed novel criteria and for which the performance is investigated in Section VI. Testing the inversion is outside the scope of the paper.

## VI. TESTS

In this section, we test the performance of our proposed approaches on PNL mixtures of speech. Indeed—and contrary to [8], [9]—we test our approach on simulations using real speech signals which can be locally sparse in the time domain, due to silence of speakers [5].

Before investigating the performance of each criterion used in each stage of the approaches, we illustrate the behaviour of the proposed methods with an example of $N = 3$ sources and $P = 2$ sensors, i.e. an underdetermined mixture. The source signals are three speech signals, which are sampled at 20 kHz, last 5 s, and contain silent parts. These sources are presented in Fig. 3.

We mixed them with the following mixing matrix:

$$A = \begin{bmatrix} 1 & 1 & 0.9 \\ -0.9 & 0.5 & 1 \end{bmatrix} \quad (25)$$

and then applied the following nonlinear mappings, proposed in [12],

$$\begin{cases} f_1(t) &= \tanh(t) + t \\ f_2(t) &= \tanh(10\,t) \end{cases} \quad (26)$$

to the resulting signals $\underline{z}(t)$. Note that the mixing matrix $A$ is close to being an absolutely degenerate matrix, and thus the configuration under consideration is challenging. Moreover, the nonlinear functions have been chosen so that they can model audio effects like *soft-clipping*. Observations are shown in Fig. 3 and one may see the strong nonlinearities in Observation $x_2(t)$.

We set the size of our temporal analysis zones to 100 samples. Mutual information is estimated using the approach in [22]. Fig. 3 shows the plot of speech sources and the obtained normalized mutual information measures. We can see that $\mathcal{I}_{\mathrm{norm}}(\underline{x})$ is close to 1 when one source is isolated. We then considered all the zones where $\epsilon_1$ and $\epsilon_2$, defined in Eq. (19), are set to 0.01 and 0.1, respectively. Fig. 4 shows two scatter plots: on the top, we provide the scatter plot of the original observations. It is clear that the sparsity assumptions needed in [8], [9] are not satisfied at all. In the bottom plot, we only show the scatter plots obtained from zones satisfying Eq. (19). Here, we can see three curves associated with nonlinearities. This thus shows the relevance of the single-source confidence measures and an easy way to improve the work of [8], [9]. We then estimated the different curves on the local scatter plots using B-spline approximations. Because the choice of the knots is data-dependent, we decided to perform a "coarse" fitting, i.e. an approximation whose knot locations and B-spline order are not necessarily optimised but that allow us to separate the curves of the functions $\phi_{ik}$ defined in Eq. (7). In the example provided here, we used the following knots

$$\forall i \in \{0, \ldots, 6\}, \quad \xi_i = -1.5 + 0.5 \cdot i, \quad (27)$$

without multiplicity order and knot-end conditions. We set the degree of the B-spline to 4, in order to obtain smooth estimates of the curves. We then obtained the B-spline coefficients $\widehat{\underline{\beta}}$ that we then clustered using K-medians. Fig. 5 shows the separated curves obtained after classification, i.e. the superposition of the local scatter plots on the zones belonging to the same cluster. Such separated curves then allowed us to estimate the inverse nonlinear mappings.

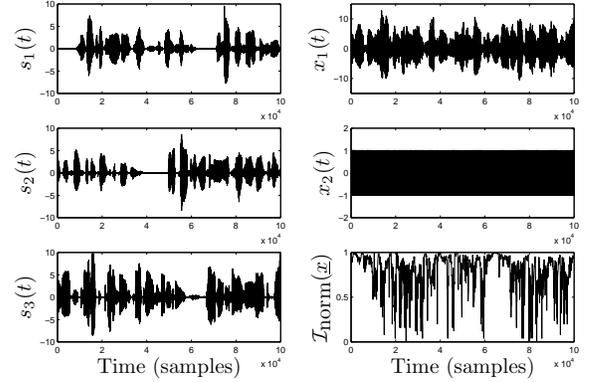

Fig. 3. Normalised mutual information measures (bottom plot on the right part of the figure) between two PNL mixtures (upper plots on the right part of the figure) of three speech sources (plotted on the left part of the figure).

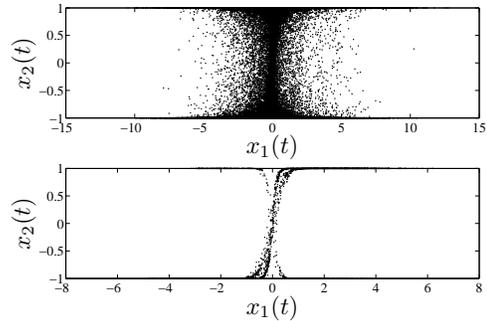

Fig. 4. Scatter plots between observations. *Top:* on the full signals. *Bottom:* on the single-source analysis zones.

This inversion can be done by e.g. applying one of the linearizing methods proposed in [8], [9]. This linearization succeeds if the nonlinear mappings are estimated with accuracy, which is the goal of this paper. We measured this accuracy by computing B-splines functions of order 4, with 20 knots. The obtained curves, drawn in Fig. 5, fit the scatter plots very well and are really close to the theoretical ones, as the mean square errors (MSEs) on the sampled curves are equal to 2.5e-4, 5.3e-5, and 2.1e-5. We now propose to characterize the performance of each proposed criterion for both finding single-source analysis zones and clustering the functional data.

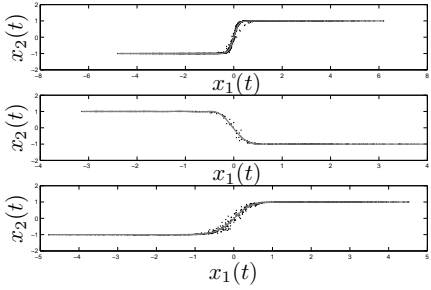

Fig. 5. Scatter plots of the separated curves after the clustering procedure (in black) and estimated nonlinear functions (in gray).

### A. Comparison of the single-source confidence measures

In this section, we test the performance of the proposed single-source confidence measures. To do so, we propose an original experiment which consists of generating linear instantaneous mixtures $\underline{z}(t)$ as defined in Eq. (1), their PNL versions $\underline{x}(t)$ defined in Eq. (2), and compute the linear and nonlinear single-source confidence measures over $\underline{z}(t)$ and $\underline{x}(t)$, respectively, in the same zones $\mathcal{T}$. These tests will allow us to derive some particular statistics of the behaviour of each considered single-source confidence measure. We hereafter denote the considered nonlinear single-source confidence measure by SSCM($\underline{x}$), and the associated linear one by SSCM($\underline{z}$). From now on we consider that a zone is selected as single-source in the linear mixtures (respectively PNL mixtures) if SSCM($\underline{z}$) > $\eta_{\underline{z}}$ (respectively SSCM($\underline{x}$) > $\eta_{\underline{x}}$), where we refer to the thresholds associated with the single-source confidence measures for PNL and LI mixtures as $\eta_{\underline{z}}$ and $\eta_{\underline{x}}$, respectively. On the contrary, when these measures are below these thresholds, we consider the zone $\mathcal{T}$ to be multiple-source.

We then define (i) TP $(\eta_{\underline{x}}, \eta_{\underline{z}})$ the true positive cases of zones $\mathcal{T}$ detected as single-source in the linear mixtures and in the PNL ones, (ii) FN $(\eta_{\underline{x}}, \eta_{\underline{z}})$, the false negative cases of zones detected as single-source in the linear mixtures but not in the PNL ones, (iii) FP $(\eta_{\underline{x}}, \eta_{\underline{z}})$ the false positive cases of zones detected as single-source in the PNL mixtures but not in the linear ones, and (iv) TN $(\eta_{\underline{x}}, \eta_{\underline{z}})$, the true negative cases where the zones $\mathcal{T}$ are not detected as single-source neither in the linear mixtures nor in the PNL ones.

From these cases, we derive the sensitivity and the specificity as:

$$\text{Sensitivity}(\eta_{\underline{x}}, \eta_{\underline{z}}) = \frac{\text{TP}(\eta_{\underline{x}}, \eta_{\underline{z}})}{\text{TP}(\eta_{\underline{x}}, \eta_{\underline{z}}) + \text{FN}(\eta_{\underline{x}}, \eta_{\underline{z}})}, \quad (28)$$

$$\text{Specificity}(\eta_{\underline{x}}, \eta_{\underline{z}}) = \frac{\text{TN}(\eta_{\underline{x}}, \eta_{\underline{z}})}{\text{FP}(\eta_{\underline{x}}, \eta_{\underline{z}}) + \text{TN}(\eta_{\underline{x}}, \eta_{\underline{z}})}. \quad (29)$$

These quantities may be analyzed as follows. The sensitivity may be seen as the probability of correctly detecting single-source zones in PNL mixtures. When this probability is low, this means that we could discard the zones $\mathcal{T}$ that would be seen as single-source by the linear measures, i.e. some single-source zones are "invisible". This might not affect the global performance of the method a lot if the total number of single-source zones is high but it might affect it if there are few single-source zones[7]. The specificity may be seen as the probability of correctly discarding multiple-source zones. If this probability is low, this means that we might detect zones $\mathcal{T}$ as single-source when they are not, thus yielding inaccurate estimation of the nonlinear functions in the next stages of the proposed approach. In our considered PNL problem, a low specificity is much more harmful than a low sensitivity and must be avoided as much as possible.

We generate 252 PNL mixtures: we consider 28 pairs of $N = 2$ speech sources, which last 5 s, include silent parts, and are sampled at 20 kHz, that we mix with the following $P \times P$ mixing matrices:

$$A = \begin{bmatrix} 1 & \lambda \\ -\lambda & 1 \end{bmatrix}, \quad (30)$$

with $\lambda = 0.9, 0.5$, and $0.1$. We then apply one of the following sets of nonlinear functions: the one in Eq. (26) tested in [12], or

$$\begin{cases} f_1(t) &= \tanh(t) + 0.1 \cdot t \\ f_2(t) &= t \end{cases} \quad (31)$$

which has been tested in [8], and

$$f_1(t) = f_2(t) = \tanh(t), \quad (32)$$

tested in [9].

The cardinality of the zones $\mathcal{T}$ is set to 100 and the temporal overlap between the zones is set to 10%, thus generating 1111 analysis zones per PNL mixtures. As in the underdetermined illustrative example described at the beginning of this section, mutual information is estimated using the approach in [22]. The eigenvalue decomposition is processed through a classical PCA [7][8]. As the locally-linear approaches described in Section III-B use a $K$-NN method for estimating the linear subspace, we vary $K$ on $\{5, 10, 15, 20\}$.

Figure 6 shows the sensitivity and the specificity for all the tested single-source confidence measures, for thresholds $\eta_{\underline{z}}$ and $\eta_{\underline{x}}$ ranging from 0 to 0.99 with a step-size of 0.01, except for the LTSA-PCA measures[9] where the thresholds range from 0.5 to 0.99. The plots may be analyzed as follows. The mutual information provides almost the same measures for both the LI and PNL mixtures, hence the symmetry which appears in the plots in Fig. 6a. Figures 6b–6e highlight different aspects of the LTSA-PCA measure. We notice that the measures are sensitive to the choice of $K$. For example, the sensitivity for high values of $\eta_{\underline{x}}$—typically $\eta_{\underline{x}} > 0.9$—decreases when $K$ increases. Similarly, the area of high values of specificity increases with $K$. Figures 6f–6i illustrate a different behaviour of the LTSA-Correlation measure. For example, these plots show that this measure is less sensitive to the choice of $K$ than the previous LTSA-PCA. In particular, the specificity is much higher for a much wider range of values of thresholds, thus showing that the risks of keeping "false" single-source zones with the LTSA-Correlation is lower than with the LTSA-PCA. However, the specificity is lower as well, thus meaning that the LTSA-Correlation discards more "real" single-source zones than the LTSA-PCA. However, and as discussed above, this last aspect is less detrimental than the previous one. To conclude, the normalized mutual information and the correlation-based confidence measure should be preferred and the eigendecomposition-based measure should be avoided. In the rest of the paper, we only use the mutual information to find single-source zones.

---

[7]Indeed, if there are few and a non-negligible part of them cannot be found, then we estimate the linear functions with an extremely few zones which might result in inaccurate estimation of the nonlinear functions.

[8]We also tested an alternative eigenvalue decomposition using a Schur decomposition, as proposed in [15]. However, we did not notice any major difference between this decomposition and the PCA-based one [7].

[9]Indeed, the LTSA-PCA measure and its linear version are based on the ratio of eigenvalues. In the case of $P = 2$ observations, the lowest eigenvalue is between 0 and 1, hence the range of the ratio which goes from 0.5 to 1.

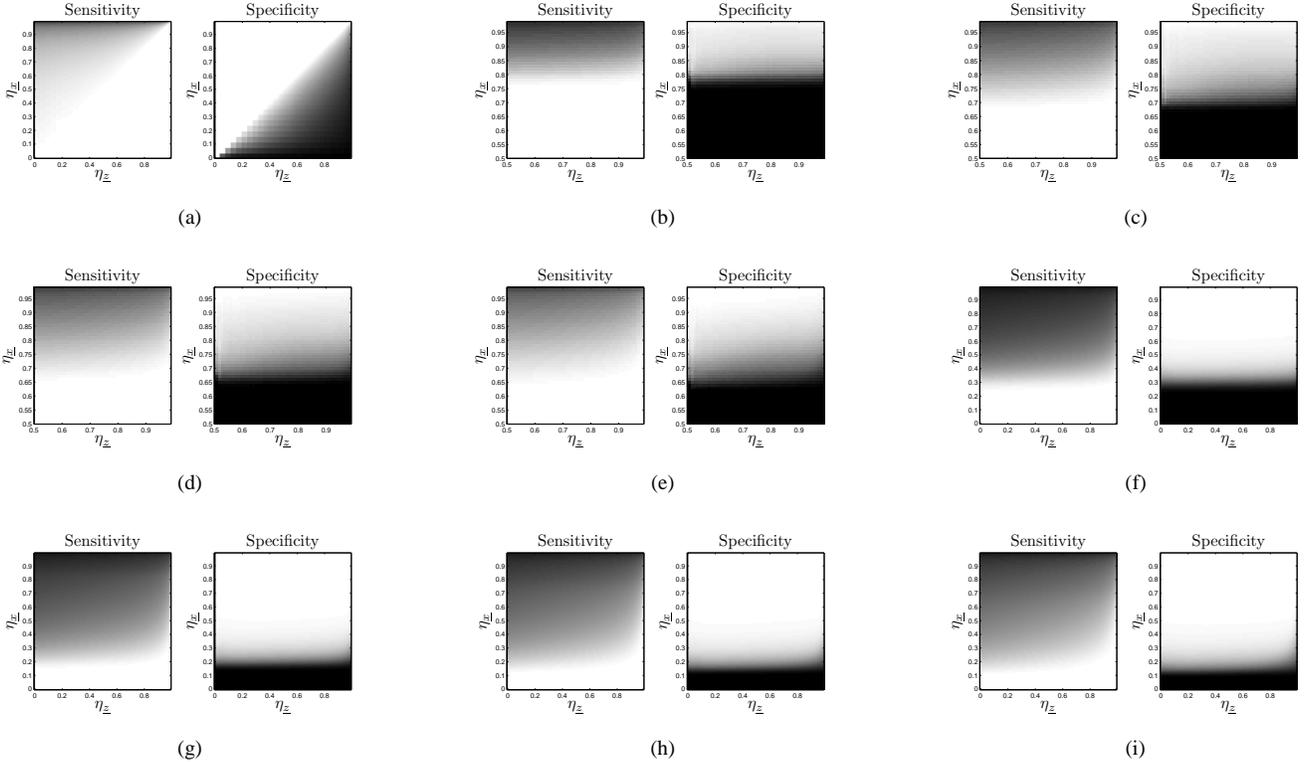

Fig. 6. Sensitivity and specificity, with respect to the value of $\eta_{\underline{z}}$ and $\eta_{\underline{x}}$, for the tested single-source confidence measures. White (respectively black) color means that the measure is equal to 1 (respectively 0). (a) Normalized mutual information. (b)–(e) LTSA-PCA with $K = 5, 10, 15$, and $20$. (f)–(i) LTSA-Correlation with $K = 5, 10, 15, 20$.

### B. Comparison of the functional data clustering approaches

We now compare the performance of the functional clustering techniques. For that, we use the same protocol as in Section VI-A. We considered the same sets of sources, the same mixing matrices and the same nonlinear functions. We selected the single-source zones by estimating the mutual information between observations. The threshold $\epsilon_1$ in Eq. (19) was set to 0.01 while the threshold $\epsilon_2$ was set[10] to 0.05. The final estimation of the nonlinear mappings is obtained by computing a B-spline of order 4 with 205 knots. Let us note that, when this last measure is quite high—say in the order of $10^{-1}$—it does not necessarily mean that the clustering stage was not successful, but that the estimation of the nonlinear mappings is not accurate enough. A better approximation may be obtained with other approaches (e.g. with another B-spline order or with kernel regression). However, this measure allows us to compare the performance obtained with the proposed clustering methods.

We aimed to measure the influence of the mixing parameters on the global quality of estimation of the nonlinear mappings, and the influence of the parameters to be in the clustering methods. The filtering clustering method uses a B-spline approximation, for which we need to fix the knot locations and the order. We proposed to test the approach with evenly-spaced knots between -1.5 and 1.5, with a step-size of 0.5 or 0.3. Additional tests—not reported here for space considerations— were performed with evenly-spaced knots ranging from -2 and 2, with a step-size of 1, 0.5, or 0.25. However, these experiments yielded to a lower performance than those reported in this paper. For each of these tests, the order of the spline varied from 3 to 6. In each experiment, when the Schoenberg-Whitney condition was not satisfied for a given single-source zone $\mathcal{T}$, this zone was discarded. The approaches using the linear approximation around zero use a $K$-NN technique. We set the value of $K$ to vary as 5, 10, 15, and 20.

---
[10]In some preliminary tests, we found that the value of this threshold had almost no effect on the global performance of the clustering methods. This was expected, as explained in Section. IV-B.

The average MSEs and their associated standard deviations are given in Table I. We first noticed that the performance of the clustering approach using B-splines highly depends on the choice of the knot locations. In particular—and this result is not evident from the table—we noticed that the MSEs can be very low with one of the three tested sets of functions but really high for another one. Even if we may get very accurate results, as we showed in [12], [13], we *never* found a set of knots and a spline order yielding very accurate estimates for all the tested nonlinear mappings. The best performance was obtained with a step between knots equal to 0.5, and a spline of order 4 or 5. However, this performance is much lower than the one obtained with the other proposed clustering methods. Indeed, the linear approximation-based techniques yield accurate estimates, except when $\gamma = 0.1$, which is consistent with the performance of other PNL-ICA methods (see [1, Ch. 14]) and the one obtained in [5] for an LI-SCA method. For the other tested mixing matrices, the accuracy of clustering and estimation of the nonlinear mappings is really good, which shows that these methods are more flexible. Moreover, the approaches do not seem to be very sensitive to the choice of $K$ as the performance does not change a lot, except when $K = 5$ where the MSEs are higher with the other tested values of $K$. This shows the relevance and the flexibility of the proposed methods.

## VII. CONCLUSION AND FUTURE WORK

In this paper, we introduced several PNL mixture identification methods which use weak sparsity assumptions in order to estimate the mixing parameters. The main novelty of the proposed method is that we combine single-source confidence measures with functional data clustering techniques. The proposed approaches thus improve on the previously proposed PNL-SCA methods which assume strong joint-sparsity assumptions and cannot be applied to speech and audio signals. We conducted several experiments showing the performance and the relevance of our approaches. In future work, we will propose an approach for inverting the nonlinearities. Indeed, here we focused on the estimation of the nonlinear functions, whose accuracy has a direct consequence on the final separation. Inversion approaches presented in [8], [9] can easily be used with our proposed approaches but we will propose an alternative to them and will compare their respective performance. Another future direction will consist of investigating sparsifying transforms well-suited to nonlinear mixtures. Indeed, while the proposed approach may be applied to speech signals, the required sparsity assumption is not met with music signals. Lastly, we will extend our approach to other nonlinear mixtures, like PNL convolutive mixtures.

## APPENDIX A
## GENERAL NONLINEAR MIXTURES

The methods we propose in this paper, for both finding single-source zones and clustering the scattered functional data, may be applied to more general mixtures than the PNL ones, as we will now see.

We assume that $N$ real source signals $\underline{s}(t) = [s_1(t), \ldots, s_N(t)]^T$ are mixed by an unknown instantaneous nonlinear mapping $\mathcal{A}$ from $\mathbb{R}^N \to \mathbb{R}^P$. Observed signals then read:
$$\underline{x}(t) = \mathcal{A}(\underline{s}(t)). \quad (33)$$
This mixing model is extremely general and it is well known [1, Ch. 14] that it cannot be solved by *only* assuming source mutual independence. Assumption 1 may now be rewritten in this new framework as follows.

*Assumption 3:* (i) The nonlinear mapping $\mathcal{A}$ is smooth, (ii) we assume we know the value of $\mathcal{A}$ for one value $\underline{u}_0 \in \mathbb{R}^N$, and in particular, without loss of generality, we assume that $\underline{u}_0 = \underline{0}$ and that
$$\mathcal{A}(\underline{0}) = 0. \quad (34)$$
Lastly, (iii) $\mathcal{A}$ may be completely estimated by its values in single-source zones $\mathcal{T}$:
$$x_i(t) = \mathcal{A}_i(\underline{s}(t)) \triangleq \mathcal{A}_{ik}(s_k(t)), \forall i \in \{1, \ldots, P\}, \quad (35)$$
where $\mathcal{A}_{ik}$ is an invertible nonlinear function from $\mathbb{R} \to \mathbb{R}$. Assumptions 3.(i) and 3.(iii) are needed in order to interpolate $\mathcal{A}$ from $\mathcal{A}_{ik}$. Assumption 3.(ii) is needed to suppress the ambiguities that may appear in the selection of single-source zones, as we faced with PNL mixtures in Section III-C. Note that Assumption 3.(iii) allows us to tackle many NL configurations, as we will now see. The PNL mixture model e.g. satisfies this assumption. The nonlinear mapping $\mathcal{A}$ may be rewritten as the composition
$$\mathcal{A} = \underline{f} \circ A, \quad (36)$$
where $\underline{f}$ and $A$ model each part of the PNL mixture, as defined in Section II. Assumption 3.(iii) also allows us to process the situation when each NL function $\mathcal{A}_i$ defined in Eq. (35) is written as a linear combination of NL functions $\mathcal{A}_{ij}$ defined from $\mathbb{R} \to \mathbb{R}$:
$$\mathcal{A}_i(\underline{s}(t)) = \sum_{j=1}^{N} \mathcal{A}_{ij}(s_j(t)). \quad (37)$$
In a general way, Assumption 3.(iii) allows us to estimate mappings $\mathcal{A}_i$ that can be inferred from the functions $\mathcal{A}_{ik}$ defined in Eq. (35). Assumption 2 may now be rewritten as:

*Assumption 4:* (i) Source signals are mutually independent and (ii) by considering *several* single-source analysis zones associated with the same source, the amplitude of the observations spans a "wide" range allowing the estimation of the NL functions $\mathcal{A}_{ik}$.

One then can see the connections between this considered problem and the PNL problem considered in the main part of this paper. If a source, say $s_k$ is isolated, then Eq. (35)

TABLE I
PERFORMANCE OF THE FUNCTIONAL CLUSTERING METHODS

| Filtering clustering [19] with K-medians [6] | | $\gamma$ | | | Linear approximation | | $\gamma$ | | |
|---|---|---|---|---|---|---|---|---|---|
| | | 0.1 | 0.5 | 0.9 | | | 0.1 | 0.5 | 0.9 |
| knot step: 0.5 | MSE | 0.1198 | 0.0364 | 0.0287 | using [7] | MSE | 0.1084 | 0.0406 | 0.0763 |
| spline order: 3 | Std. | 0.4539 | 0.2033 | 0.1757 | $K=5$ | Std. | 0.5587 | 0.1689 | 0.4036 |
| knot step: 0.5 | MSE | 0.1243 | 0.0079 | 0.0010 | using [7] | MSE | 0.1002 | 0.0003 | 0.0005 |
| spline order: 4 | Std. | 0.5038 | 0.0642 | 0.0040 | $K=10$ | Std. | 0.5704 | 0.0011 | 0.0022 |
| knot step: 0.5 | MSE | 0.0777 | 0.0083 | 0.0005 | using [7] | MSE | 0.1270 | 0.0004 | 0.0003 |
| spline order: 5 | Std. | 0.3766 | 0.0845 | 0.0021 | $K=15$ | Std. | 0.6864 | 0.0012 | 0.0011 |
| knot step: 0.5 | MSE | 0.6310 | 2.1020 | 5.5983 | using [7] | MSE | 0.1141 | 0.0004 | 0.0003 |
| spline order: 6 | Std. | 1.8711 | 4.2003 | 12.7026 | $K=20$ | Std. | 0.5931 | 0.0013 | 0.0011 |
| knot step: 0.3 | MSE | 1.0470 | 0.2497 | 0.1592 | using [5] | MSE | 0.1148 | 0.0524 | 0.0730 |
| spline order: 3 | Std. | 2.8847 | 0.5025 | 0.3810 | $K=5$ | Std. | 0.5564 | 0.2377 | 0.3380 |
| knot step: 0.3 | MSE | 0.3813 | 0.1840 | 0.1790 | using [5] | MSE | 0.0777 | 0.0004 | 0.0027 |
| spline order: 4 | Std. | 1.0729 | 0.3692 | 0.4804 | $K=10$ | Std. | 0.3760 | 0.0012 | 0.0261 |
| knot step: 0.3 | MSE | 0.1720 | 0.0977 | 0.1426 | using [5] | MSE | 0.1275 | 0.0004 | 0.0003 |
| spline order: 5 | Std. | 0.3546 | 0.2868 | 0.5043 | $K=15$ | Std. | 0.6881 | 0.0012 | 0.0011 |
| knot step: 0.3 | MSE | 0.2406 | 0.1509 | 0.0639 | using [5] | MSE | 0.1400 | 0.0004 | 0.0003 |
| spline order: 6 | Std. | 1.032 | 0.8957 | 0.2863 | $K=20$ | Std. | 0.7113 | 0.0013 | 0.0011 |

holds and we then obtain, assuming that $\mathcal{A}_{ik}$ is invertible,

$$s_k(t) = \mathcal{A}_{ik}^{-1}(x_i(t)) \quad \forall i \in \{1, \ldots, P\}, \forall t \in \mathcal{T}. \quad (38)$$

We thus have the following relationship between observations $x_1$ and $x_i$, for all $t \in \mathcal{T}$:

$$x_i(t) = \mathcal{A}_{ik}\left(\mathcal{A}_{1k}^{-1}(x_1(t))\right) = \phi_{ik}(x_1(t)), \quad (39)$$

where the functions $\phi_{ik}$ are defined as:

$$\phi_{ik}(u) = \mathcal{A}_{ik}\left(\mathcal{A}_{1k}^{-1}(u)\right). \quad (40)$$

Estimating single-source zones and clustering the scattered functional data in this framework can be done with the same approach that we defined for PNL mixtures. The main difficulty then consists of inverting the nonlinear functions, which is not in the scope of this paper.

To demonstrate the validity of our extension, we here consider a toy example: we generate $N = 2$ Gaussian noise sources containing $10^5$ samples, for which we set the first and last 2000 samples to 0, in the first and second source respectively. We then mixed these sources by following the mixture model defined in Eq. (37), with the following nonlinear functions:

$$\mathcal{A}_{11}(t) = \mathcal{A}_{22}(t) = \tanh(t), \quad \mathcal{A}_{12}(t) = \mathcal{A}_{21}(t) = t. \quad (41)$$

We then used the same values as in Section VI-B to set our parameters, with the above optimal parameters for the clustering approaches: the filtering method was using B-splines of order 4, with evenly spaced knots ranging from -1.5 to 1.5 with a step-size of 0.5. The LTSA-based approaches were using $K = 15$ neighbors. Nonlinear functions were lastly approximated with B-splines of order 4, with 20 knots. The experiment was conducted 10 times and the MSE between the theoretical and estimated nonlinear functions were averaged. The three methods yielded almost the same performance: the filtering approach provided a mean MSE equal to 0.0104 while both LTSA-based methods provided a MSE equal to 0.0110.

ACKNOWLEDGMENT

This work is funded by the Marie Curie IAPP "AVID MODE" grant within the European Commission's FP7.

REFERENCES

[1] P. Comon and C. Jutten, *Handbook of blind source separation. Independent component analysis and applications*. Academic press, 2010
[2] O. Yilmaz and S. Rickard, "Blind separation of speech mixtures via time-frequency masking", *IEEE Transactions on Signal Processing*, Vol. 52, No. 7, pp. 1830–1847, 2004.
[3] F. Abrard and Y. Deville, "A time-frequency blind signal separation method applicable to underdetermined mixtures of dependent sources", *Signal Processing*, Vol. 85, No. 7, pp. 1389–1403, 2005.
[4] Y. Deville, M. Puigt and B. Albouy, "Time-frequency blind signal separation: extended methods, performance evaluation for speech sources," in *Proc. of IJCNN 2004*, Vol. 1, pp. 255–260, Budapest, Hungary, 2004.
[5] Y. Deville and M. Puigt, "Temporal and time-frequency correlation-based blind source separation methods. Part I: determined and underdetermined linear instantaneous mixtures", *Signal Processing*, Vol. 87, No. 3, pp. 374–407, Mar. 2007.
[6] M. Puigt and Y. Deville, "Iterative-shift cluster-based time-frequency BSS for fractional-time-delay mixtures, " in *Proc. of ICA 2009*, LNCS, Vol. 5441, pp. 306–313, Paraty, Brazil, 2009.
[7] S. Arberet, R. Gribonval, and F. Bimbot, "A robust method to count and locate audio sources in a multichannel underdetermined mixture", *IEEE Transactions on Signal Processing*, Vol. 58, No. 1, Jan. 2010.
[8] F. Theis and S. Amari, "Postnonlinear overcomplete blind source separation using sparse sources," in *Proc. of ICA 2004*, LNCS, Vol. 3195, pp. 718–725, Granada, Spain, 2004.
[9] S. van Vaerenbergh and I. Santamaría, "A spectral clustering approach to underdetermined postnonlinear blind source separation of sparse sources", *IEEE Transactions on Neural Networks*, Vol. 17, No. 3, pp. 811–814, 2006.
[10] Y. Deville and S. Hosseini, "Blind identification and separation methods for linear-quadratic mixtures and/or linearly independent non-stationary signals," in *Proc. of ISSPA 2007*, Sharjah, UAE, Feb, 12–15, 2007.
[11] L. T. Duarte and C. Jutten, "A nonlinear source separation approach for the Nicolsky-Eisenman model," in *Proc. of EUSIPCO 2008*, Lausanne, Switzerland, 2008.

[12] M. Puigt, A. Griffin, and A. Mouchtaris, "Post-nonlinear speech mixture identification using single-source temporal zones & curve clustering," in *Proc. of EUSIPCO 2011*, Barcelona, Spain, 2011.

[13] M. Puigt, A. Griffin, and A. Mouchtaris, "Nonlinear blind mixture identification using local source sparsity and functional data clustering," to appear in *Proc. of SAM 2012*, Hoboken, NJ, USA, 2012.

[14] A. Dionisio, R. Menezes, and D. A. Mendes, "Mutual information: a measure of dependency for nonlinear time series", *Physica A: Statistical Mechanics and its Applications*, Vol. 344, No. 1–2, pp. 326–329, 2004.

[15] L.J.P. van der Maaten, E.O. Postma, and H.J. van den Herik, "Dimensionality reduction: a comparative review", *Tilburg University Technical Report*, 2009.

[16] C. Févotte and C. Doncarli, "Two contributions to blind source separation using time-frequency distributions", *IEEE Signal Processing Letters*, Vol. 11, No. 3, pp. 386–389, 2004.

[17] J.O. Ramsay and B.W. Silverman, *Functional Data Analysis*. Springer-Verlag, 2005.

[18] H.P Krieger, P. Kröger, and A. Zimek, "Clustering high-dimensional data: A survey on subspace clustering, pattern-based clustering, and correlation clustering", *ACM Trans. Knowl. Discov. Data*, Vol. 3, No. 1, pp. 1–58, 2009.

[19] C. Abraham, P. A. Cornillon, E. Matzner-Løber, and N. Molinari, "Unsupervised curve clustering using B-splines", *Scandinavian Journal of Statistics*, Vol. 30, No. 3, pp. 581–595, Sep. 2003.

[20] H.B. Curry and I.J. Schoenberg, "On Pólya frequency functions IV. The fundemental spline functions and their limits", *Journal d'Analyse Mathématiques*, Vol. 17, pp. 71–107, 1966.

[21] I.J. Schoenberg, A. Whitney, "On Pólya frequency functions III. The positivity of translation determinants with an application to the interpolation problem by spline curves", Trans. Amer. Math. Soc, Vol. 74. pp. 246-259, 1953.

[22] G.A. Darbellay and I. Vajda, "Estimation of the information by an adaptive partitioning of the observation space", *IEEE Transactions on Information Theory*, Vol. 45, No. 4, pp. 1315-1321, May 1999.

**Anthony Griffin** Anthony Griffin received his PhD in Electrical & Electronic Engineering from the University of Canterbury in Christchurch, New Zealand in 2000. He then spent three years programming DSPs for 4RF, a Wellington-based company selling digital microwave radios. He subsequently moved to Industrial Research Limited—also based in Wellington—focussing on signal processing for audio signals and wireless communications. In 2007, he joined the Institute of Computer Science, Foundation for Research and Technology-Hellas (FORTH-ICS), Heraklion, Greece as a Marie Curie Fellow, where he is working on audio signal processing, wireless sensor networks and compressed sensing. He also occasionally teaches a postgraduate course in Applied DSP at the University of Crete.

**Matthieu Puigt** Matthieu Puigt was born in Perpignan, France, in 1980. He first studied Mathematics at the Université de Perpignan where he graduated both the Bachelor in Pure Mathematics in 2001 and the Master's degree in Mathematical Engineering in 2002. He then received the DEA in Signal, Image Processing, and Acoustics, from the Université Paul Sabatier Toulouse 3, Toulouse, France, in 2003, and his Ph.D. in Signal Processing from the Université de Toulouse in 2007. From 2007 to 2009 he was a Postdoctoral Lecturer at the Université Paul Sabatier Toulouse 3 and the Laboratoire d'Astrophysique de Toulouse-Tarbes. From September 2009 to June 2010, he held an Assistant Professor position at the University for Information Science and Technology, in Ohrid, Republic of Macedonia (FYROM). Since August 2010, he is a Marie Curie postdoctoral fellow at the Institute of Computer Science of the Foundation for Research and Technology – Hellas (FORTH-ICS). Matthieu Puigt's current research interests include signal processing, sparsity, unsupervised classification, and especially Blind Source Separation methods and their applications to Acoustics and Astrophysics. He has authored or co-authored more than 15 publications in journal or conference proceedings and has served as a reviewer for several scientific journals and international conferences in these areas.

**Athanasios Mouchtaris** Athanasios Mouchtaris received the Diploma degree in Electrical Engineering from Aristotle University of Thessaloniki, in 1997 and the M.S. and Ph.D. degrees from the University of Southern California (USC), Los Angeles, in 1999 and 2003, respectively. From 2003 to 2004 he was a Postdoctoral Researcher in the Electrical and Systems Engineering Department of the University of Pennsylvania, Philadelphia. From 2004 to 2007 he was a Postdoctoral Researcher in the Institute of Computer Science of the Foundation for Research and Technology Hellas (FORTH-ICS), Heraklion, Crete, and a Visiting Professor in the Computer Science Department of the University of Crete. Since 2007 he has been an Assistant Professor in the Computer Science Department of the University of Crete, and an Associate Researcher in FORTH-ICS. His research interests include signal processing for immersive audio environments, spatial audio rendering, multichannel audio modeling, speech synthesis with emphasis on voice conversion, and speech enhancement. He has authored or co-authored more than 50 publications in various journal and conference proceedings in these areas. He has received grants from the European Commission and the Greek General Secretariat of Research and Technology.